  \documentstyle[12pt]{article}
  
\catcode`@=11
\def\secteqno{\@addtoreset{equation}{section}%
\def\theequation{\thesection.\arabic{equation}}}
\catcode`@=12
\secteqno
\newcommand{\be}{\begin{equation}}
\newcommand{\ee}{\end{equation}}
\newcommand{\bea}{\begin{eqnarray}}
\newcommand{\eea}{\end{eqnarray}}
\newcommand{\bref}[1]{(\ref{#1})}
\newcommand{\ep}{\epsilon} 
\newcommand{\T}{\theta} 
   
\newcommand{\A}{\alpha} \newcommand{\B}{\beta}
\newcommand{\G}{\Gamma} \newcommand{\D}{\delta}
         \newcommand{\lam}{\lambda}
\newcommand{\r}{\rho}           \newcommand{\s}{\sigma}

\newcommand{\h}{\eta}           
           
\newcommand{\W}{\Omega}

\def\pa{\partial}

\def\CL{{\cal L}}

\def\CF{{\cal F}}
\def\Tb{{\overline\theta}}

\def\EE{e^{\frac{\phi}{2}}(E+\chi B)}

\def\nb{\nabla}

\newcommand{\nn}{\nonumber}

\def\t{\tilde}

\def\l{{\ell}}

\def\gg{{\tau_3}}

\newcommand{\slPi}{/ {\hskip-0.27cm{\Pi}}}

\newcommand{\slbz}{/ {\hskip-0.27cm{\beta}}_0}

\begin{document}
          \hfill TOHO-FP-9860  

	  \hfill YITP-98-24

\vskip 20mm 
\begin{center} 
{\bf \Large  Self-duality in Super D3-brane Action} 
\vskip 10mm
{\large Yuji\ Igarashi, Katsumi\ Itoh,$^a$\footnote{Permanent Address: Faculty of Education, Niigata University, Niigata 950-21,
Japan} and Kiyoshi\ Kamimura,$^b$}\par 

\medskip
{\it 
Faculty of Education, Niigata University, Niigata 950-21, Japan\\
$^a$ Yukawa Institute for Theoretical Physics, Kyoto University\\Kyoto
606-01, Japan\\
$^b$ Department of Physics, Toho University, \ Funabashi\ 274-8510, Japan\\
}

\medskip
\date{\today}
\end{center}
\vskip 10mm
\begin{abstract}

We establish self-duality of super D3-brane theory as an exact symmetry
of the action both in the Lagrangian and Hamiltonian formalism.  In the
Lagrangian formalism, the action is shown to satisfy the Gaillard-Zumino
condition. This algebraic relation is recognized in our previous
paper to be a necessary and sufficient condition for generic action of
U(1) gauge field strength coupled with gravity and matters to be
self-dual. For the super D3-brane action, SO(2) duality transformation of a
world-volume gauge field should be associated with SO(2)
rotation of fermionic brane coordinates in N=2 SUSY multiplet. This
SO(2) duality symmetry is lifted to SL(2,R) symmetry in the presence of
a dilaton and an axion background fields.  In the canonical formalism,
we show that the duality rotation is described by a canonical
transformation, and the Hamiltonian of the D3-brane action is invariant under
the transformation.

\end{abstract}


\newpage
\setcounter{page}{1}
\setcounter{footnote}{0}
\parskip=7pt 


\section{Introduction} 
\indent 

SL(2,Z) symmetry has been recognized to play an important role in
understanding S duality.  The type IIB D-branes appear in multiplets
under this SL(2,Z) symmetry.  Among others the D3-brane has a special
position to be a singlet, thus often referred to be self-dual.  Our main
purpose in this paper is to show the self-duality of the super D3-brane
based on our earlier works\cite{IIKK},\cite{IIK}.  Actually, there have
appeared some papers claiming this self-duality as Tseytlin's for
bosonic action\cite{Tseytlin} and Aganagic et al.'s for super
action\cite{Aganagic2}; however those works are based on semi-classical
treatments.  We would like to emphasize that our result to be reported
here does not depend on any semi-classical approximations.

The precise statement of the self-duality is that the super D3
action\footnote{Hereafter the D3-brane means {\it super} D3-brane if not
stated otherwise.} is invariant under a combined transformation of
vector duality on the world-volume gauge field and the SL(2,Z)
transformation of the external supergravity backgrounds.  Since the
world-volume is four-dimensional, the vector duality for D3-brane is
nothing but the well-known electric-magnetic duality.  Usually the
electric-magnetic duality is regarded as a symmetry at the level of
equations of motion (EOM)\footnote{Self-duality of bosonic D3-brane was
discussed in ref.\cite{GreenGurperle} as a symmetry of the EOM.}.  This makes
the self-duality of the D3 {\it action} highly non-trivial.

In the following we discuss our strategy to show the self-duality. The
proof is given both in the Lagrangian and the Hamiltonian approaches.

In the Lagrangian formalism, our proof is based on an alternative
description of electric-magnetic duality rotation in terms of the gauge
potential\cite{DeserTeitelboim}, which was given in our previous
paper\cite{IIK}. (This rotation is referred to as {\it
A-transformation}, while the conventional one for the field strength as
{\it F-transformation}.) It enables us to formulate the duality as a
symmetry of generic actions of U(1) gauge field strength coupled with
gravity and matters. In the same paper we also emphasized that a
necessary and sufficient condition for the invariance\footnote{Strictly
speaking, the actions are not exactly invariant but pseudo-invariant,
which means that the actions remain invariant only up to surface terms.} 
of the action may be expressed as an algebraic relation, the
Gaillard-Zumino (GZ) condition
\cite{GaillardZumino},\cite{GaillardZumino2},\cite{GibbonsRasheed}.
Here we show that the D3 action obeys this condition, thereby
establishing its self-duality as an exact symmetry without resort to any
semi-classical approximations.  It should be stressed that, in order for
the action to be invariant, SO(2) duality rotation of the gauge field
should be associated with SO(2) rotation of the fermionic brane
coordinates, while the bosonic coordinates yet remain unchanged.

The proof for self-duality is also discussed in the Hamiltonian
formalism, where the duality symmetry is realized as an invariance of
the Hamiltonian.  Based on general analysis of type IIB super D-branes
given in ref.\cite{Kamimura}, we investigate transformation properties of
the constraints in the D3 action: the bosonic constraints are shown to
be invariant under the duality transformation, and fermionic ones to
transform by the SO(2) rotation mentioned above. These give a proof of
the invariance of the Hamiltonian for the D3 action.  The covariance of
the fermionic constraint is a consequence of the transformation
properties of the fermionic coordinates, which is determined to satisfy
the GZ condition from the viewpoint of the Lagrangian formalism.

We also confirm the idea that the vector duality transformations can be
essentially identified with canonical transformations. Based on a
canonical analysis of the D-string\cite{HaK}, we gave in a previous
paper\cite{IIKK} the canonical transformation that relates the D-string
action with the type IIB superstring action.  We construct here, as a
natural extension of that work, the canonical transformation to generate
the A-transformation for the D3 action.

To write down D3 action explicitly, we need an integrated expression of
the Wess-Zumino term.  It was given by Cederwall et al.\cite{Cederwall}
and by Kamimura and Hatsuda\cite{Kamimura}. Two Lagrangian densities
differ only by a total derivative and they essentially give the same
action.\footnote{Here expected is some cohomology argument similar to
the one given in \cite{Azcarraga}} Here we will take the action of
ref.\cite{Cederwall} below. The action contains a dilaton and an axion
as scalar SUGRA backgrounds, which lift the duality symmetry from SO(2)
to SL(2,R)\cite{GibbonsRasheed2}: they become the variables
parametrizing the coset space SL(2,R)/SO(2), and give a non-linear
realization of the SL(2,R) symmetry.  For simplicity, we assume these
scalars to be constant fields, though extension to the on-shell SUGRA
multiplet is possible\cite{Cederwall}.

This paper is organized as follows: the next section describes the
duality condition for generic action of an interacting U(1) gauge
field strength. In section 3, we show that the D3 action obeys the
SO(2) duality condition.  The SO(2) duality is shown to be lifted to
the SL(2,R) duality in the presence of the dilaton and the axion.  The
proof of invariance or covariance of the constraints in the
Hamiltonian formalism is given in section 4. The final section is
devoted to summary and discussion.  Here some comments will also be
made on other approaches to implement the duality symmetry at the
action level and further to relate it to a possible non-perturbative
definition of string theory.

\section{The Gaillard-Zumino condition}
\indent

We begin with a brief summary of the GZ condition\cite{GaillardZumino},\cite{GaillardZumino2},\cite{GibbonsRasheed}, which is discussed in
detail in ref.\cite{IIK}. Consider a generic Lagrangian density
$\CL(F_{\mu\nu}, g_{\mu\nu}, \Phi^A)=\sqrt{-g} L(F_{\mu\nu}, g_{\mu\nu},
\Phi^A)$ in D=4, which depends on a U(1) gauge field strength
$F_{\mu\nu}$, metric $g_{\mu\nu}$, and matter fields $\Phi^A$.  The
constitutive relation is given by
\bea
\t K^{\mu\nu}~=~\frac{\pa L}{\pa F_{\mu\nu}},~~~~~~~~~~
\frac{\pa F_{\A\B}}{\pa F_{\mu\nu}}~=(\D^\mu_\A~\D^\nu_\B~-~
\D^\mu_\B~\D^\nu_\A),
\label{defKt}
\eea
where the Hodge dual components\footnote{We use the following
convention: $\eta^{\mu\nu\r\s}$ denotes the covariantly constant
anti-symmetric tensor with indices raised and lowered using the metric
$g_{\mu\nu}$ whose signature is $(-+++)$. We also use the tensor
densities $\ep^{\mu\nu\r\s}$ and $\ep_{\mu\nu\r\s}$ with weight $-1$ and
$1$.  They are defined by $\ep^{\mu\nu\r\s} =\sqrt{-g}\eta^{\mu\nu\r\s}$
and $\eta_{\mu\nu\r\s} =\sqrt{-g} \ep_{\mu\nu\r\s}$ with $g= \det
g_{\mu\nu}$, normalized as $\ep^{0123}~=~-\ep_{0123}~=~1$.} for
the anti-symmetric tensor $K_{\mu\nu}$ are defined by
\bea 
\t K_{\mu\nu}~=~\frac{1}{2} \eta_{\mu\nu}^{~\r\s}~K_{\r\s},~~~~~~
{\tilde {\t K}}_{\mu\nu}~ =~-~K_{\mu\nu}.  
\eea
Gaillard and Zumino considered an infinitesimal duality transformation 
which consists of the most
general linear transformation on $F$ and $K$, and a
transformation of matter fields,
\bea
\D\pmatrix{F\cr K}~=~\pmatrix{\alpha&\beta\cr \gamma&\delta}\pmatrix{F\cr K},~~~~~~
\D\Phi^A~=~\xi^A(\Phi),~~~~~~\D g_{\mu\nu}=0,
\label{DT}
\eea
and required invariance of stationary surfaces of the system under
\bref{DT}.  It was shown\cite{GaillardZumino} that \\ 
(1) the F-transformation in \bref{DT} is an element of
SL(2,R) given by $\delta =-\alpha$; \\
(2) the Lagrangian should transform as
\bea \D L~=~\frac14(\gamma~F~\t F~+~\beta~K~\t K).  
\label{Dual} 
\eea 
As to be seen later, the non-compact SL(2,R) duality is possible only
when there are scalar fields in the theory. In their absence, the
relevant duality group becomes SO(2): the transformation described by
the compact maximal subgroup, $U(1)\sim SO(2)$, where the parameters
satisfy the conditions $\alpha=-\delta=0,~\beta=-\gamma\equiv\lam$. The
SO(2) transformation is given by
\bea \D F~=~\lam~K,~~~~~\D K~=~-~\lam~F.  \label{DFK} \eea
Since the Lagrangian changes by
\bea
\D L~=~\frac12~\frac{\pa L}{\pa F_{\mu\nu}}\D F_{\mu\nu}~
+~\frac{\pa L}{\pa \Phi^A}\D \Phi^A~=~
\frac{\lam }{2}~\t K^{\mu\nu}~K_{\mu\nu}~+~\D_\Phi L~,
\label{DL}
\eea
the duality condition \bref{Dual} reduces to 
\bea
\frac{\lam}{4}(~F~\t F~+~K~\t K)~+~\D_\Phi L~=~0,
\label{u1duality}
\eea
where $F~\t F= F_{\mu\nu}{\t F}^{\mu\nu}$. 

As shown in \cite{IIK}, eq.\bref{u1duality} is the crucial condition for
a generic action for an interacting U(1) field to be self-dual: this
algebraic relation is not an on-shell relation but sensible even for
off-shell fields; furthermore, in the formulation based on the
A-transformation, the GZ condition \bref{u1duality} ensures the duality
as a symmetry of the action.

A comment is in order.  In the GZ condition, the first two terms may
be obtained with the definition for $K$ once an action is specified.
So the question of the duality reduced to a problem to find an
appropriate matter transformation so that the GZ condition is
satisfied. In this sense, it may be regarded as a condition on the
matter transformations.


\section{D3 action and the Gaillard-Zumino condition}

In this section we first consider D3 action without scalar SUGRA
backgrounds and show that it satisfies the SO(2) duality condition
\bref{u1duality}. Let $X^M$ be bosonic brane coordinates in D=10 flat
target space $~(M=0,...,9)$, and $\T_{A \alpha}$ be its fermionic
partners described by the Majorana-Weyl spinor with spinor index
$\alpha$ and N=2 SUSY index $A$.  We shall use the same conventions for
the Dirac matrices as those given in ref.\cite{Aganagic}.  These indices for
spinors are suppressed below.  The D3 action for the brane coordinates
$(X,\T)$ and world-volume gauge field $A_{\mu}$ is required to have
the kappa symmetry and N=2 SUSY. It takes the form
\bea
S&=&\int~d^4\s~\CL^{DBI}~+~\int d^4 \sigma ~\CL^{WZ}
\label{DBILs}
\eea
where
\bea
\CL^{DBI}&=&-~\sqrt{-\det(G_{\mu\nu}+\CF_{\mu\nu})},
~~~~~~G_{\mu\nu}~=~\Pi_{\mu}^M\Pi_{\nu M},~~~
\nn\\
\CF_{\mu\nu}&=&
\pa_{[\mu}A_{\nu]}~+~\W_{\mu\nu}^3,~~~~~
\W_{\mu\nu}^j~=~\Tb~\hat \slPi_{[\mu}~\tau_j~\pa_{\nu]}\T~~~(j=1,3) .
\eea
The Pauli matrices $\tau_i$ act on N=2 SUSY indices.
The basic one-form is defined by  
\bea
\Pi^M &\equiv& dX^{M} ~+~{\bar \theta}~{\Gamma}^{M} ~d{\theta}~\equiv~d{\sigma}^{\mu}~\Pi_\mu^{M},~~~~~~~~
\Pi^M_{\mu}~=~{\partial}_{\mu} X^M~-~{\bar \theta}~{\Gamma}^M~
{\partial}_{\mu}\theta.
\eea
and
\bea
{\hat \Pi}^{M}~=~\Pi^M~-~\frac12~{\bar \theta}
{\Gamma}^{M}~d{\theta}~=~dX^M~+~\frac12~{\bar \theta}{\Gamma}^{M}~d\theta.
\eea
For the Wess-Zumino (WZ) action, we take the one given in ref.\cite{Cederwall}.  Using differential forms, it is
given by the 2-form $\cal F$, a pullback of Ramond-Ramond 2-form $C^{(2)}$
and a 4-form $C^{(4)}$:
\bea
{L^{WZ}}&=&C^{(2)}~\CF~+~C^{(4)}~
\label{wzactC}
\\
C^{(2)}&=&\Tb~\hat\slPi~\tau_1d\T~=~\W_1,
\label{C2}
\\
C^{(4)}&=&\Xi~-~\frac12~\W_1~\W_3, 
\label{C4}
\eea
where
\bea
\Xi&=&\frac16~\Tb~\slPi^3~\tau_3\tau_1~d\T
\nn\\
&-&\frac1{12}~\Tb~(\slPi^2\slbz~+~\slPi\slbz\slPi~+~\slbz\slPi^2)
\tau_3\tau_1~d\T
\nn\\
&+&\frac1{18}~\Tb~(\slPi\slbz^2~+~\slbz\slPi\slbz~+~\slbz^2\slPi)
\tau_3\tau_1~d\T
\nn\\
&-&\frac1{12}~\Tb~\slPi~\tau_{[1}~d\T~\Tb~\slbz~\tau_{3]}~d\T
\nn\\
&-&\frac1{24}~\T~\slbz^3~\tau_3\tau_1~d\T,~~~~~~~~~~~~~~~~~~~~~~
(\B_0~\equiv~\Tb~\G~d\T).
\label{Q}
\eea

Now let us see whether the GZ condition is satisfied for the above
action.  First of all, we calculate the first two terms of the
condition.  From the definition in \bref{defKt} the $\t K$ is obtained
as,
\bea
\t K^{\mu\nu}&=&\frac{\pa L}{\pa F_{\mu\nu}}
\nn\\
&=&\frac{\sqrt{-G}}{\sqrt{-G_\CF}}(\CF^{\nu\mu}~+~{\cal T}~\t \CF^{\mu\nu})~
+~\t C^{(2)\mu\nu},
\label{defKtsd3}
\eea
where use has been made of the determinant formula for the
four-by-four matrix;
\bea
G_{\CF}\equiv \det(G+\CF)~=~G\left(1~+~\frac12\CF^{\mu\nu}\CF_{\mu\nu}~+~
{\cal T}^2 \right),~~~~~~~
{\cal T}~\equiv~\frac14 \CF_{\mu\nu}\t \CF^{\mu\nu}.
\eea
Taking the Hodge dual of \bref{defKtsd3}, we find the $K$ as,
\bea
K_{\mu\nu}
&=&-~\frac12\eta_{\mu\nu\r\s}~\t K^{\r\s}
=~\frac{\sqrt{-G}}{\sqrt{-G_\CF}}(\t \CF_{\mu\nu}~+~{\cal T}~ \CF_{\mu\nu})~
+~C^{(2)}_{\mu\nu}.
\label{defKd3s}
\eea
The last terms in \bref{defKtsd3} and \bref{defKd3s} arise from the
first term in the WZ term in (3.2),
\bea
C^{(2)}~\CF~=~\frac{1}{4}d^4\s~\ep^{\mu\nu\r\s}~C^{(2)}_{\mu\nu}~\CF_{\r\s}.
\eea
The product of \bref{defKtsd3} and \bref{defKd3s} gives
\bea
K_{\mu\nu}~\t K^{\mu\nu}&+&F_{\mu\nu}~\t F^{\mu\nu}
\nn\\
&=&-2\t F^{\mu\nu}\W_{\mu\nu}^3~-~\W_{\mu\nu}^3\t\W^{\mu\nu}_3~
+~2~\t K^{\mu\nu}C^{(2)}_{\mu\nu}~-~\t C^{(2)\mu\nu}~C^{(2)}_{\mu\nu}.
\label{310}
\eea

It may be appropriate to make a few remarks on the bosonic truncation of
the D3 action, $L_B$.  Obviously the r.h.s. of \bref{310} vanishes in
this case.
Substituting the relation \bref{310} into \bref{Dual} with
$\beta=-\gamma=\lambda$ for $SO(2)$, we obtain the variation of the
Lagrangian as $\D L = - \frac{\lambda}{2}~F~{\tilde F}$: the bosonic DBI
Lagrangian density transforms into a total
derivative.\footnote{Tseytlin\cite{Tseytlin} discussed the
pseudo-invariance of this bosonic action for the flat metric case.}
Eq.\bref{310} also implies from the GZ condition that $\delta_{X} L_{B}
=0$: so $\delta X =0$ is a right assignment for the matter transformation.

Let us turn to the supersymmetric case and discuss the matter
contribution in the GZ condition.  It is used to find an appropriate
transformation for the matter fields in such a way that it makes the
action invariant under the dual transformation.  For our present case of
the D3 brane, we will find that the following transformation for the
matter fields, $X$ and $\theta$, suites our purpose:
\bea
\D\T&=&\lam\frac{i\tau_2}{2}~\T,~~~~~~~~\D X~=~0, 
\label{delsT}
\eea
which gives
\bea \D\Pi^M_{\mu}=\delta G_{\mu \nu}= \D \Xi=0,~~~~~~
\D\W_{\mu\nu}^3=-~\lam~\W^{1}_{\mu\nu},~~~~~~\D
\W^{1}_{\mu\nu}~=~\lam\W_{\mu\nu}^3.  
\label{metricinv}
\eea 
Note that the Majorana-Weyl fermions $(\T_1,~\T_2)$ and $(\W^1,~\W^3)$
transform as SO(2) doublets. Presently we will find the invariance of
$\Xi$ under duality rotation is crucial to satisfy the duality condition.

We turn to the variation of the total Lagrangian in \bref{DBILs}
with respect to the matter transformation,
\bea
\D_{\Phi} L&=& \D_{\theta} L =\frac12\frac{\pa L}{\pa\CF_{\mu\nu}}\D\W_{\mu\nu}^3~+~\frac12
\t\CF^{\mu\nu}\D C^{(2)}_{\mu\nu}~+~\D \t C^{(4)},
\label{311}
\eea 
where ${\t C^{(4)}}$ is the Hodge dual of ${C^{(4)}}$:
$C^{(4)}=d^4\s\frac1{4!}\ep^{\mu\nu\r\s}C^{(4)}_{\mu\nu\r\s}\equiv
d^4\s \sqrt{-G}\t C^{(4)}$.  $C^{(2)}=\W_{1}$ and the invariance of
$\Xi$ give rise to a relation of the differential forms
\bea
\D \Xi =~\D~[~\frac{1}{2}~C^{(2)}~\W_3~+~ C^{(4)}~]~=
\frac{\lam}{2}(~-(C^{(2)})^2~+~(\W_3)^2~)~+~\D_\T C^{(4)}~
=~0.
\label{consd32}
\eea

Combining the results in \bref{310}, \bref{311} and \bref{consd32}, we 
find 
\bea
& &\frac{\lam}{4}(~F~\t F~+~K~\t K)~+~\D_{\Phi} L\nn\\
& &~~~=\frac{\lam}{4}(~-C^{(2)}_{\mu\nu}~\t C^{(2)\mu\nu}~
+~\W_{\mu\nu}^3\t\W^{\mu\nu}_3~)~
+~\D \t C^{(4)}~=0~~.
\label{consd31}
\eea
Therefore, the duality condition is satisfied.

It has been recognized that the SO(2) duality may be lifted to the
SL(2,R) duality by introducing a dilaton $\phi$ and an axion
$\chi$\cite{GibbonsRasheed2}, \cite{GaillardZumino}. They are assumed to
be constant background fields. According to the general method, one
defines a new Lagrangian using the D3 Lagrangian $L(F, X, \T)$ which
obeys the SO(2) duality
\bea
\hat L(F,~X,~\T;~\phi,~\chi)~=~L(e^{-\phi/2}F,~X,~\T)~-~\frac14\chi~ F~\t F.
\label{hatL}
\eea
If one introduces $\hat F=e^{-\phi/2}F$ and $\hat K$ by taking the dual
of $(-)\pa L(\hat F,X,\T)/\pa \hat F$, the background dependence is
absorbed in the rescaled variable $(\hat F, \hat K)$. These are related
with the background dependent $(F, K)$ by
\bea
\pmatrix{F\cr K}~=~V~\pmatrix{\hat F\cr \hat K},~~~~~~~~
V~=~e^{\phi/2}~\pmatrix{1 & 0 \cr -\chi & e^{-\phi}}.
\label{FKV}
\eea
Here $V$ is a non-linear realization of SL(2,R)/SO(2) transforming as
\bea
V~~~\rightarrow~~~{\Lambda}~V~O(\Lambda)^{-1}.~~~~~
\label{transV}
\eea
Here $\Lambda$ is a global SL(2,R) matrix
\bea
\Lambda&=&\pmatrix{a&b\cr c&d}~\in~SL(2,R),~~~~~ad-bc~=~1
\eea
and $O(\Lambda)$ is an SO(2) transformation 
\bea
O(\Lambda)^{-1}~=~\pmatrix{\cos\lam&\sin\lam\cr-\sin\lam&\cos\lam}~\in~SO(2).
\eea 
This ``compensating'' transformation is induced so that the form of
$V$ is unchanged: 
\bea
\cos\lam~=~\frac{a-b\chi}{\sqrt{(a-b\chi)^2+b^2~e^{-2\phi}}},~~~~~~~
\sin\lam~=~\frac{-b~e^{-\phi}}{\sqrt{(a-b\chi)^2+b^2~e^{-2\phi}}}.  
\eea
This procedure enables us to make the SO(2) dual theory discussed above
to an SL(2,R) dual theory.

\vskip 6mm

\section{Hamiltonian formalism of D3 action}
\indent

In this section, we give a proof of self-duality of the D3 action in the
Hamiltonian formalism, using a general analysis of
constraints\cite{Kamimura} for super D-brane actions in type IIB theory.  Let us
include dilaton and axion, $\phi$ and $\chi$, from the beginning.  From
\bref{hatL} the action is given by
\bea
S&=&\int~d^4\s~\CL^{DBI}~+~\int~L^{WZ}~-~\int~\frac12\chi~F^2
=~\int~d^4\s~{\cal L}^{total},
\eea
where the 2-form component $\CF_{\mu\nu}$ appeared in $\CL^{DBI}$ and
$L^{WZ}$ is replaced by
\bea
\CF_{\mu\nu}&=&e^{-\frac{\phi}{2}}~F_{\mu\nu}~+~\W_{\mu\nu}^3.
\eea
Let $(X^M,~P_M),~(\T,~\pi_{\T})$, and $(A_{\mu},~E^{\mu})$ be
canonically conjugate pairs of the phase space variables, and define
the three-dimensional anti-symmetric tensor by $\ep^{ijk}=\ep^{0ijk}$.
We will soon find it useful to define the following new variables,
\bea
{\cal B}^i~=~\frac12\ep^{ijk}\CF_{jk}
~=~e^{-\frac{\phi}{2}}~B^i~+~\frac12\ep^{ijk}\W_{jk}^3,~~~
B^i~=~~\frac12\ep^{ijk}F_{jk},~~~(i,j,k=1,2,3) 
\label{defBt}
\eea
and 
\bea
{\cal P}_M&\equiv&\frac{\pa\CL^{DBI}}{\pa \Pi^M_0}~=~P_M~-~
\EE^i~\frac{\pa\CF_{0i}(\Pi,\dot\T)}{\pa \Pi^M_0}~-~\frac{\pa\CL^{WZ}(\Pi,\CF,\dot\T)}
{\pa \Pi^M_0},
\label{defpt}\\
{\cal E}^i&\equiv&\frac{\pa\CL^{DBI}}{\pa \CF_{0i}}~=~
\EE^i~-~\frac{\pa\CL^{WZ}(\Pi,\CF,\dot\T)}{\pa \CF_{0i}}.
\label{defEts}
\eea
In the last equation use has been made of the equation:
$E^i=\partial {\cal L}^{total}/~\partial F_{0i}$.
We find constraints of the system to be given by: \\
(1) the $U(1)$ constraints,
\bea
E^0~=~0,~~~~~~~~~~\pa_i~E^i~=~0;
\eea
(2) the $p+1$ diffeomorphism constraints,
\bea
\varphi_i&\equiv&{\cal P}\cdot\Pi_i~+~{\cal E}^j~\CF_{ij}~=~
{\cal P}\cdot\Pi_i~+~\ep_{ijk}~{\cal E}^j~{\cal B} ^k~=~0,~~~~(i=1,2,3)
\nn\\
\varphi_0&\equiv&\frac12~[~{\cal P}^2~+~\gamma~+~\gamma_{ij}~({\cal E}^i~{\cal E}^j~+~
{\cal B}^i~{\cal B}^j)~]~=~0;
\label{defH}
\eea
(3) the fermionic constraints,
\bea 
\psi&\equiv&~\pi_{\theta}~-~P_M~\frac{\pa\Pi^M_0}{\pa \dot\T}~-~\EE^i~
\frac{\pa\CF_{0i}(\Pi,\dot\T)}{\pa \dot\T}~-~
\frac{\pa\CL^{WZ}(\Pi,\CF,\dot\T)}{\pa \dot \T}~=~0.
\label{defF} 
\eea
Here $\gamma_{ij}$ is the spatial part of the induced metric and
$\gamma$ is its determinant.

We now show the invariance of the bosonic constraints ($\varphi_{0}$
and $\varphi_{i}$) and the covariance of the fermionic constraints
($\psi$) under SL(2,R) transformation of $(B,E)$ and $(\phi,\chi)$
associated with the SO(2) rotation of the fermionic fields $\T$.
To this end, we rewrite \bref{defBt} and \bref{defEts} as
\bea
\pmatrix{{\cal B} \cr{\cal E}}~=~V^{-1}
\pmatrix{ B\cr E}~+~
\pmatrix{\W^3\cr-\W^1},~~~~~~~~~
{(\W^\l)}^i~\equiv~\ep^{ijk}\Tb~\hat\slPi_j\tau_\l~\pa_k\T,
\label{49}
\eea
where $V$ is an SL(2,R)/SO(2) matrix given in \bref{FKV}.  Under
SL(2,R) transformation, $\pmatrix{{\cal B} \cr{\cal E}}$ rotates into
$O(\Lambda)\pmatrix{{\cal B} \cr{\cal E}}$ as an SO(2) vector: each
element of the first term in \bref{49} transforms by
\bea
\pmatrix{ B\cr E}~&\rightarrow&~\Lambda~\pmatrix{ B\cr E},~~~~~ 
\Lambda\in~SL(2,R),
\\
V^{-1}~=~e^{\frac{\phi}{2}}\pmatrix{e^{{-\phi}}&0\cr\chi& 1}~&\rightarrow
&~O(\Lambda)~V^{-1}~\Lambda^{-1},~~~~~~~~O(\Lambda)\in SO(2).
\eea
Likewise, the second term transforms by
\bea
\pmatrix{\W^3\cr-\W^1}~\rightarrow~O(\Lambda)~\pmatrix{\W^3\cr-\W^1}
\label{w1w3}
\eea
under the $\tau_2$ rotation of spinors $\T$,
\bea
\T \rightarrow {\cal O}(\Lambda)~\T.
\label{delsT1}
\eea
Here ${\cal O}(\Lambda)$ corresponds to the fundamental (spin 1/2) 
representation of SO(2).

Next we consider $\cal P$ given by
\bea
{\cal P}_M 
&=&
P_M~-~\left({\cal E}^i \frac{\pa \CF_{0i}(\Pi,\dot\T)}{\pa \Pi^M_0}+{\cal B}^i
\frac{\pa C^{(2)}_{0i}(\Pi,\dot\T)}{\pa \Pi^M_0}\right)\nn\\
&{}&~~~~
-~
\left(\frac12\ep^{ijk}C^{(2)}_{jk} \frac{\pa\CF_{0i}(\Pi,\dot\T)}{\pa \Pi^M_0}+
\frac{\pa}{\pa \Pi^M_0} \sqrt{-G}\t C^{(4)}(\Pi,\dot\T)\right)
\label{tp}.
\eea
We will see that each term in the above expression, and thus $\cal P$
itself, is invariant under the duality transformation.  The invariance
of $X^M$ implies that $P_M$, the conjugate variable, is also
invariant.\footnote{We will make a more accurate statement on the
invariance of $P_M$ at the end of this section.}  The second term on the
r.h.s. of \bref{tp} may be rewritten as
\bea
\left({\cal E}^i \frac{\pa\CF_{0i}}{\pa \Pi^M_0}+
\frac12\ep_{ijk}{\cal B} ^i\frac{\pa {C^{(2)}}^{jk}}{\pa \Pi^M_0}\right)
&=&
{\cal E}^i \Tb\G_M\gg\pa_i\T~+~{\cal B} ^i~\Tb\G_M\tau_1\pa_i\T~
\nn\\
&=& ({\cal B} ^i,~{\cal E}^i)\pmatrix{\Tb\G_M\tau_1\pa_i\T\cr \Tb\G_M\gg\pa_i\T}~=~
\Tb\G_M\t\tau^i_{0}\pa_i\T,
\label{e1}
\eea
where
\bea
\t\tau^i_{0}~=~{\cal E}^i \gg~+~{\cal B} ^i~\tau_1.
\eea
When sandwiched between $\T$'s, $~\t\tau^a_{0}$ is invariant under
simultaneous rotation of $\T$ and $({\cal B} ,{\cal E})$.  So the
second term is invariant.  In terms of the differential form, the last
term on the r.h.s. of \bref{defpt} is expressed as
\bea
{C^{(2)}}^{0i} \frac{\pa \CF_{0i}}{\pa \Pi^M_0}&+&
\frac{\pa}{\pa \Pi^M_0} {\sqrt{-G} \t C^{(4)}} 
\nn\\
&\rightarrow&\frac12[-({\bar \theta}~{\Gamma}_{M}{\tau}_{[3} d{\theta})({\bar \theta}~\hat\slPi\tau_{1]}~d{\theta})-
({\bar \theta}~{\Gamma}_M\tau_{(3} d{\theta})({\bar \theta}~\hat\slPi\tau_{1)}~d{\theta})]_{\bf 3}~+~
\left[\frac{\pa {C^{(4)}}}{\pa \Pi^M_0}\right]_{\bf 3}
\nn\\
&=&\frac12[-({\bar \theta}~{\Gamma}_M\tau_{[3} d{\theta})({\bar \theta}~\hat\slPi\tau_{1]}~d{\theta})]_{\bf 3}~
+~\left[\frac{\pa}{\pa \Pi^M_0}\left({C^{(4)}}~
+~\frac12\W_1\W_3\right)\right]_{\bf 3},
\label{3form}
\eea
where $[~~]_{\bf 3}$ denotes a spatial 3-form coefficient of
$[~~]$. In the last expression we observe that two terms are invariant
separately: the first is written with an anti-symmetrization of
$\tau_1$ and $\tau_3$ and is invariant under the $\theta$ rotation; the
second term is obviously related to the invariant quantity
$\Xi~=~[{C^{(4)}}~+~\frac12\W_1\W_3]$.  This completes our proof of the
invariance of $\cal P$.

In the diffeomorphism constraints, ${\cal B} $ and ${\cal E}$ appear
only in SO(2) invariant combinations. Therefore, one concludes that
$\varphi_{0}$ and $\varphi_{i}$ are invariant.

We may see the covariance of the fermionic constraint in parallel
with the above discussion on $\cal P$. The constraint is expressed as 
\bea
\psi&=&\pi_{\theta}~+~P_M~({\bar \theta}~{\Gamma}^M)~
-~\left({\cal E}^i \frac{\pa\CF_{0i}}{\pa \dot {\theta}}+
\frac12\ep_{ijk}{\cal B} ^i\frac{\pa {C^{(2)}}^{jk}}{\pa \dot{\theta}}\right)
\nn\\
&{}&~~~-
\left({C^{(2)}}^{0i} \frac{\pa\CF_{0i}}{\pa \dot{\theta}}+
\frac{\pa {\sqrt{-G}{\t C}^{(4)}}}{\pa \dot{\theta}}\right),
\label{F}
\eea
where one finds
\bea
\left({\cal E}^i \frac{\pa\CF_{0i}}{\pa \dot {\theta}}+
\frac12\ep_{ijk}{\cal B} ^i\frac{\pa {C^{(2)}}^{jk}}{\pa \dot{\theta}}\right)~=~
\frac12{\bar \theta}{\Gamma}\t\tau^i_{0}\pa_i{\theta}~\cdot{\bar \theta}{\Gamma}~-~{\bar \theta}{\Gamma}\t\tau^i_{0}\hat\Pi_i,
\label{e3}
\eea
and 
\bea
\left({C^{(2)}}^{0i} \frac{\pa\CF_{0i}}{\pa \dot{\theta}}+
\frac{\pa}{\pa \dot{\theta}} {\sqrt{-G}{\t C}^{(4)}}\right)
&\rightarrow& \Bigl[-\frac12({\bar \theta}~\hat\slPi\tau_{[1}~d{\theta})
\bigl\{({\bar \theta}~{\Gamma}_M\tau_{3]} d{\theta})~\frac12{\bar \theta}{\Gamma}^M~-~{\bar \theta}~\hat\slPi\tau_{3]}~\bigr\}
\nn\\ 
&{}&~~~~+~
\frac{\pa \Xi}{\pa \dot {\theta}}\Bigr]_{\bf 3}.
\label{e42}
\eea
These expressions and the fact that $\pi_{\theta}$ transforms as
$\pi_{\theta}\rightarrow  \pi_{\theta}{\cal O}(\Lambda)^{T}$
imply that the
fermionic constraint $\psi$ transforms covariantly:
$\psi \rightarrow  \psi{\cal O}(\Lambda)^{T}.$

We now show that the A-duality transformation is described as a
canonical transformation, using our results\cite{IIK} on the curved
space extension of the A-transformation.  It should be remarked that, in
the case at hand, the intrinsic D=3 metric $g_{ij}$ used in ref.\cite{IIK}
is replaced by the induced metric $\gamma_{ij}$ expressed in terms of
brane coordinates, $(X,~\T)$: the invariance of D=3 metric puts a
non-trivial condition on the induced metric, which is satisfied for the
present case as shown in \bref{metricinv}.

For any function on the phase space $R(p,q)$, the transformation

\bea 
\D R~=~-i[R,~{\cal W}],
\label{canotra}
\eea
is defined via the generator
\bea
{\cal W}&=&\lambda \int d^3\s {\sqrt
      \gamma}~\left[\frac{1}{2}~\frac{E^i}{\sqrt \gamma}
{D^{-1}_{i\l}}~\frac{E^\l}{\sqrt \gamma}~+~\frac{1}{2}~ A_i~{D^{ij}} A_j
~+~ \pi_{\T} 
\frac{i\tau_{2}}{2} \T \right].
\label{canogene}
\eea
Here we have used the following operators for D=3 covariant formulation:
$D^{-1}_{i\l}$ is a tensor operator acting on a vector,
\bea
D^{-1}_{i\l}\equiv(\t\Delta^{-1})_i^{~k}\nb^j\h_{jk\l}
~=~\h_{ijk}\nb^k(\t\Delta^{-1})^j_{~\l},
\label{defDI}
\eea
where $\h_{jk\l}~=~\ep_{jk\l}~{\sqrt \gamma}$ is the covariantly constant
anti-symmetric tensor.  It is the inverse of  
\bea
D^{jk}~=~\h^{j\l k}~\nb_\l~=~\nb_\l~\h^{j\l k},
\label{defD}
\eea
in a projected space 
\bea
D^{-1}_{im}~D^{mk}&=&O_i^{~k}(\nb),~~~~~~~~
D^{im}~D^{-1}_{mk}~=~O^i_{~k}(\nb),
\nn\\
O_i^{~k}(\nb)&=&\D_i^{~k}~-~\nb_i(\Delta^{-1})\nb^k.
\label{FM1}
\eea
Note that the operator $O_i^{~k}(\nb)$ projects out any longitudinal
component defined with the covariant derivative $\nb_i$.

The curved space extension of the Laplacian operator
$(\t\Delta)_{j}^{~i}$, which maps a vector $T_i$ into a vector
$(\t\Delta)_{j}^{~i}T_{i}$, is given by

\bea
(\t\Delta)_{j}^{~i}~=~\Delta \delta_{j}^{~i}-R_{j}^{~i},
\label{MLap}
\eea
where $\Delta~=~\nb^j \nb_j$ and $R_{j}^{~i}$ is the Ricci tensor. We
assume that boundary conditions can be arranged so that the Laplacian
operator has no non-trivial kernel, and its inverse,
$(\t\Delta^{-1})_i^{~j}$, is well-defined.

One finds that $\cal W$ in \bref{canogene} generates the desired
A-duality transformation for the gauge field as well as the SO(2)
rotation of $\T$:
\bea
\D A_\l~&=&~
\lam~(\t\Delta^{-1})_\l^{~k}\nb^j\ep_{jkm}~E^{m}
~=~D^{-1}_{\l m}~\left(\lam~\frac{E^{m}}{\sqrt \gamma} \right)
\\
\D E^i&=&-{\lam}~\ep^{ijk}~\pa_j A_k;
\\
\D \T &=& \lambda \frac{i\tau_{2}}{2} \T~~~~~~\D \pi_{\T} = - \pi_{\T} \lambda \frac{i\tau_{2}}{2};
\label{canotra2}\\
\D X &=& 0,~~~~~\D P = 0.
\label{canotra3}
\eea
It follows from the above expressions that
\bea
\D B^i = \lambda E^{i}_{\perp},~~~~~~~\delta E^{i}_{\perp}=
-\lambda B^i,
\eea
where 
\bea 
E^{k}_\perp = E^{k}~-~ {\sqrt
\gamma}\nb^k(\Delta^{-1})~\left(\nb_m~\frac{E^{m}}{\sqrt \gamma}\right).
\label{defZL}  
\eea
This describes duality  
exchange between the electric and magnetic fields.  

In \bref{canotra2} and \bref{canotra3}, although not described
explicitly, there appear some additional terms in the transformations of
momenta, $(\D P,~\D \pi_{\T} )$. It is because the relevant metric is
the induced one given via $(X, \T)$, and the metric dependent term in
\bref{canogene} generates new contributions to $(\D P,~\D \pi_{\T} )
$. These terms, however, are shown to be proportional to the Gauss law
constraint, $\pa_i E^{i} =0$, and therefore do not affect the
transformation rule \bref{canotra} on the constraint surfaces.

In summary we have shown that the constraint equations 
of D3 action are invariant or covariant under:
\begin{enumerate}
\renewcommand{\labelenumi}{\arabic{enumi})}
\item linear SL(2,R) transformation of 
$\pmatrix{B\cr  E}~\rightarrow~\Lambda\pmatrix{B\cr  E}$;
\item rotation of $({\theta},~\pi_{\theta})$ by $({\cal
O}(\Lambda){\theta},~\pi_{\theta} {\cal O}(\Lambda)^T)$, induced by the
SL(2,R) transformation;
\item non-linear transformation of the backgrounds $\phi$ and $\chi$ as
\bref{transV}.
\end {enumerate}
Note that $(X, ~P)$ are left invariant (up to the Gauss law constraint).

This completes the proof of the invariance of the Hamiltonian, the
self-duality in the canonical formalism.  It is worth mentioning that
the duality transformation does not commute with global SUSY
transformation. This is suggested by the fact that Majorana-Weyl spinors
$\T$ transform under the duality transformation, while the bosonic
counterpart $X^M$ is left invariant. The SUSY charge $Q$ undergoes the
same transformation as $\pi_{\T}$ under the duality rotation: $Q
\rightarrow Q~{\cal O}(\Lambda)^{T}$.  In this connection, note that the
NS-NS two form $B^{(2)}=-\Omega ^3$ and R-R two form $C^{(2)}=\Omega^1$
mixed as an SO(2) vector.

\section{Summary and Discussion}

In D=4 spaces, irrespective of being target space or world-volume, the
vector duality transformation is special in the sense that it is nothing
but the electric-magnetic duality rotation. Gaillard and Zumino showed
that the maximal group of the duality transformation allowed for an
interacting gauge field strength is the SL(2,R). They also found the
duality condition: the Lagrangian needs to transform in a particular way
under the duality rotation in order for EOM to remain invariant or
covariant. It turns out, however, that the GZ condition is more than
that: this algebraic relation is the necessary and sufficient condition
for invariance of the action, and may serve therefore a guiding
principle of constructing D=4 actions of U(1) gauge field strength
coupled with gravity and matters in string and field theories. Our proof
of the self-duality of the D3 action may be the first non-trivial
application of this idea.

Obviously, the existence of the criterion for duality symmetry such as
the GZ condition in the Lagrangian formalism is only possible in D=4
theories including D3-action.  As for the other D-branes in the
effective action approach to string and M-theory, we believe that the
Hamiltonian formalism suits better for establishing exact symmetries or
relations. Actually, we showed in a previous article\cite{IIKK} the
canonical equivalence between D-string action and IIB string action.
The previous work and the present one for D3-action strongly support the
idea that the vector dualities in type IIB theory can be understood as
canonical transformations. We expect furthermore that the relationships
between the D-brane actions in type IIA theory and the dimensionally
reduced M-brane actions may be understood similarly in the canonical
formalism.

It would be appropriate to make some comments on other approaches to
implement the duality symmetry at the action level.  In \cite{Bengtsson}
the DBI action is reformulated in a duality manifest way by introducing
another world-volume gauge field. This approach is an extension of the
Schwarz-Sen model\cite{SchwarzSen} with two gauge fields, and its
covariant version, the PST model\cite{Tonin}.  
Possible relations among higher dimensional theories, two gauge field
formulations and a manifestly dual invariant formulation of string
(effective) theories are extremely interesting, though there must be
many points to be clarified to find them in concrete.  In particular, to
formulate the {\it supersymmetric} version of two gauge models, the
knowledge of fermion transformations would be crucial.  It is very
interesting to imagine that there is a condition in two gauge models, an
extension of the GZ condition, which puts some restrictions on matter
transformations and helps us to find yet unknown supersymmetric
extensions.  \footnote{A super D3-brane action with two gauge fields was
given in ref.\cite{Westerberg} in a different context(see also
ref.\cite{Townsend}). It would be interesting to investigate how duality
symmetry is realized there.}

Our A-transformation approach has its own drawbacks: non-locality and
the sacrifice of the manifest D=4 covariance.  The two gauge field
formulation and its extensions have been introduced to overcome these
difficulties.  An extension of our argument given in this paper
may be extended to those approaches.  However we did not take those
point of views because of the following reasons: our present approach is
enough to show the self-duality; and we believe that there are much more
to be done to figure out a real relation of those approaches to the
non-perturbative string theory.

\medskip\noindent
{\bf Acknowledgements}\par \medskip\par
K.I. would like to thank to the Yukawa Institute for its kind
hospitality extended to him.


\end{document}